\documentclass{ws-Preprint} 

\begin{document}
\title{\Large Temperature and Density Conditions for Alpha Clustering in
Excited Self-Conjugate Nuclei}
\author{
Bernard Borderie $^{1}$, Adriana Raduta $^{2}$, Enrico De
Filippo $^{3}$,  Elena Geraci $^{4,3}$, Nicolas Le Neindre $^{5}$,
Giuseppe Cardella $^{3}$, Gaetano Lanzalone $^{6,7}$, Ivano Lombardo
$^{3}$, Olivier Lopez $^{5}$, Concettina Maiolino $^{6}$, Angelo Pagano
$^{3}$, Massimo Papa $^{3}$, Sara Pirrone $^{3}$, Francesca Rizzo
$^{4,6}$ and Paolo Russotto $^{6}$
\vspace{6pt}
}
\address{%
$^{1}$ Universit\'e Paris-Saclay, CNRS/IN2P3, IJCLab, 91405 Orsay, France\\
$^{2}$ National Institute for Physics and Nuclear Engineering
(IFIN-HH), 077125 Bucharest, Romania\\
$^{3}$ INFN, Sezione di Catania, 95125 Catania, Italy\\
$^{4}$ Dipartimento di Fisica e Astronomia ``Ettore Majorana'',
Universit\`a di Catania, 95123 Catania, Italy\\
$^{5}$ Normandie Univ, ENSICAEN, UNICAEN, CNRS/IN2P3, LPC, 14050 Caen, France\\
$^{6}$ INFN, Laboratori Nazionali del Sud, 95125 Catania, Italy\\
$^{7}$ Facolt\`a di Ingegneria e Architectura, Universit\`a Kore,
94100 Enna, Italy\\
\vspace{6pt}
(corresponding author: bernard.borderie@ijclab.in2p3.fr)
}
\maketitle
\abstracts{\underline{Abstract:}
Starting from experimental studies on alpha-clustering in excited
self-conjugate nuclei (from $^{16}$O to $^{28}$Si), temperature and
density conditions for such a clustering are determined. 
Measured temperatures have been found in the range of 5.5 - 6.0 MeV
whereas density values of 0.3 - 0.4 times the saturation density 
are deduced, i.e., 0.046 to 0.062 $fm^{-3}$.  
Such a density domain is also
predicted by constrained self-consistent mean field calculations.
These results constitute a benchmark for alpha clustering from 
self-conjugate nuclei in relation to descriptions of stellar 
evolution and supernovae.}

\section{Introduction}
The knowledge of the composition of warm nuclear matter at low density is
of paramount importance for a better understanding of
the description of the core-collapse of supernovae as well as for the
formation and static properties of proto-neutron
stars~[\cite{Fischer14,Arcones08}].
 In this context, cluster formation is one of the 
fundamental aspects with, in particular, the role of $\alpha$-particles
which are predicted to be present due to the instability of nuclear
matter against cluster formation~[\cite{Rop98,Bey00,Hor06,Sam09,Typ10}]. 
Related to this, the formation of $\alpha$-particle 
clustering from excited expanding self-conjugate
nuclei was also revealed in two different constrained self-consistent mean
field calculations~[\cite{girod_prl2013,ebran_2014,ebran_2020}].
On the experimental side,
cluster formation and their in-medium effects
have been probed by heavy-ion
experiments~[\cite{Qin12,Hem15,Pais20PRL,Pais20JPG}] and
alpha-clustering was observed in excited expanding self-conjugate
nuclei~[\cite{Bor16,Bor17,Bor15}].
The aim of the present paper is to give a benchmark for
the temperature and density needed to observe alpha-clustering in
self-conjugate nuclei. The information is derived from 
experimental data (temperature, freeze-out volume/density) and 
densities compared with theoretical expectations. 
The paper is organized as follows. In section 2 the experimental
context and the event selection will be presented. Section 3 is
dedicated to alpha-particle clustering (evidence and deduced
temperature-density information). 
Finally, in section 3 density information from the theoretical side is
discussed.  
\section{Experiment and event selection}
The experiment was performed at INFN,
Laboratori Nazionali del Sud in Catania, Italy.
The chosen experimental strategy was to use the reaction $^{40}Ca$ + $^{12}C$ 
at an incident energy (25 MeV per nucleon) high enough to produce some
hot expanding fragmentation products~[\cite{Ann90,Mor85}]
associated to a high
granularity large solid angle particle array to precisely reconstruct
the directions of the velocity vectors.
The beam impinging on a thin carbon target (320 $\mu$g/cm$^2$)
was delivered by the Superconducting Cyclotron and the
charged reaction products were detected by the CHIMERA 4$\pi$
multi-detector~[\cite{chimera}]. The beam intensity was kept around
$10^7$ ions/s to avoid pile-up events and random coincidences, which
is mandatory for high multiplicity studies. 
CHIMERA consists of 1192 telescopes ($\Delta$E silicon detectors
200-300 $\mu$m thick and CsI(Tl) stopping
detectors) mounted on 35 rings covering 94\% of the solid angle,
with very high granularity at forward angles.
Details on A and Z identifications and on the quality of
energy calibrations can be found
in Refs.~[\cite{chimera,Ald02,Len02,Bor16}].  
One can just underline that careful identifications and selections were 
allowed by a complete exclusive detection in $A$ and $Z$ of all reaction products 
and by the excellent forward granularity of CHIMERA; 
energy resolution was better than 1\% for silicon detectors and varies
between 1.0 and 2.5\% for alpha particles stopped in CsI(Tl) crystals. 

As a first step in our event selection procedure, we want to exclude
poorly-measured events. Without making any
hypothesis about the physics of the studied reaction one can measure
the total detected charge $Z_{tot}$ (neutrons are not measured).
Due to their cross-sections  and to the
geometrical efficiency of CHIMERA, the well detected reaction
products correspond to projectile
fragmentation (PF)~[\cite{Ann90,Mor85}] with $Z_{tot}$ = 19-20
(target reaction products not
detected) and to incomplete/complete fusion
with $Z_{tot}$ = 21-26~[\cite{Eud14}].
At this stage we can have the first indication of the multiplicity
of $\alpha$-particles, $M_{\alpha}$, emitted per event for 
well identified mechanisms 
($Z_{tot} \geq$ 19 - see figure 1 from Ref.~\cite{Bor16}). 
$M_{\alpha}$ extends up to thirteen, which means a
deexcitation of the total system into $\alpha$-particles. Moreover, a
reasonable number of events exhibit $M_{\alpha}$ values up to about 6-7.

The goal is now to tentatively isolate, in the detected
events, reaction products emitting
$\alpha$-particles only. Refs.~[\cite{Mor85,Fuc94}]
 have shown that, at incident
energies close to ours, $^{20}$Ne or $^{32}$S PF is dominated by alpha-conjugate
reaction products. Based on this, and expecting the same for $^{40}$Ca, we
restrict our selection to completely detected PF events ($Z_{tot}$ = 20)
composed of one projectile fragment and $\alpha$-particles.
Charge conservation imposes $Z_{frag}$ = 20 - 2$M_{\alpha}$.
An example of the mass distribution of the single fragment can be seen 
in Ref.~\cite{Bor16}.

After this double selection, the question is: from which emission
source are the $\alpha$-particles emitted?
Several possibilities are present and further
selections must be made before
restricting our study to alpha-sources emitting exclusively the $M_{\alpha}$
observed (called $N\alpha$ sources in what follows).
Possibilities that we must examine are the following:

 I) considering the incident energy of the reaction and the forward
  focusing of reaction products, it is important to identify
  the possible presence of preequilibrium (PE) $\alpha$-particles in
  our selected PF events. With the hypothesis that all the 
  $\alpha$-particles are emitted from their center-of-mass reference
  frame, we noted an energy distribution which resembles a thermal one
  with the presence of a high energy
  tail starting at 40 MeV, which signs 
  PE emission (see figure 3 from Ref.~\cite{Bor16}). 
  To prevent errors on alpha emitter properties, it is necessary
  to remove events in which such PE emission can be present;
  an upper energy limit of 40 MeV found irrespective of
  $M_{\alpha}$ was imposed to the $\alpha$-particle energy. At
  this stage 6.9 to 9.2\% of events were excluded.
  
 II) $\alpha$-particles can be emitted from deexcitation of  PF events via
  unbound states of $^{12}$C, $^{16}$O, $^{20}$Ne  and not directly from
  excited expanding $N\alpha$ sources. 
  We want, for instance, to exclude from the selection an event
  composed of two fragments ($^{24}$Mg and $^{12}$C*) and one
  $\alpha$-particle finally producing one single fragment ($^{24}$Mg)
  and four $\alpha$-particles.
  Multi-particle correlation
  functions~[\cite{Cha95,Lis91}] were used to identify
  unbound states $\alpha$-particle
  emitters and to exclude a small percentage of events (1.6-3.9\%).

III) it must be verified that the fragments associated with
  $M_{\alpha}$ are not the evaporation residues of excited $Ca$
  projectiles that have emitted sequentially $\alpha$-particles only.  

As far as the two first items are concerned the effect was to
suppress from 8.5 to 12.8\% 
of previously selected events; more details can be found 
in Ref.~\cite{Bor16}. The last item will be discussed in the
following section.
\section{Experimental results}
\subsection{Evidence for alpha-particle clustering}
Before discussing different possible deexcitation mechanisms 
involved in the retained events, information on the projectile 
fragmentation mechanism is needed. Global features of PF events 
are reproduced by a model of stochastic transfers~[\cite{Tass91}].
The main characteristics for primary events with $Z_{tot}$=20 are the
following: i) excitation energy
 extends up to about 200 MeV, which 
allows the large excitation energy domain
(20-150 MeV) measured for $N\alpha$ sources when associated to a single
fragment (see Figure~\ref{fig:EXCIT_T_b}); ii) angular 
momenta extend up to 24 $\hbar$, which gives
an upper spin limit for Ca projectiles or $N\alpha$ sources.
We recall that excitation energies of $N\alpha$ sources
are equal to the sum of kinetic energies of particles in the $N\alpha$
reference frame plus the reaction Q value.
\begin{figure}[!ht]
\centering
\includegraphics[width=0.43\textwidth]{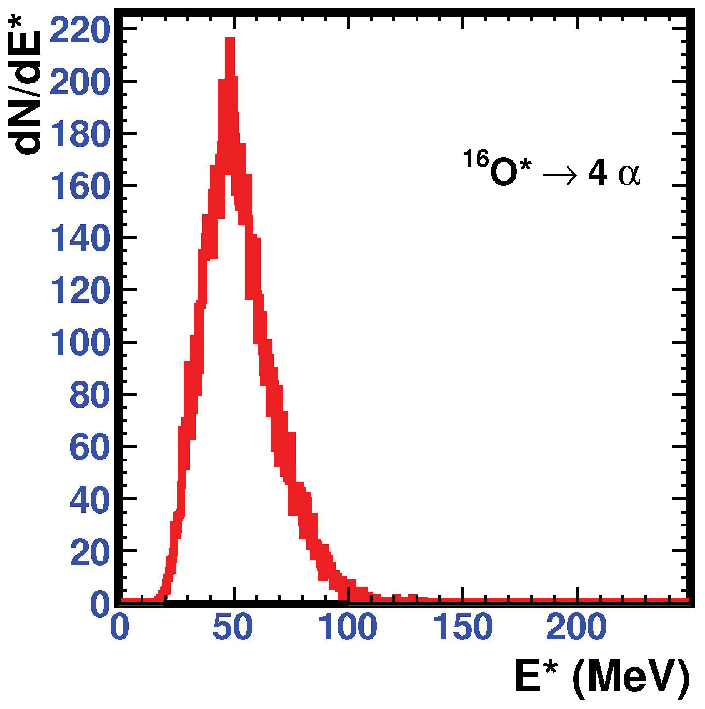}
\includegraphics[width=0.43\textwidth]{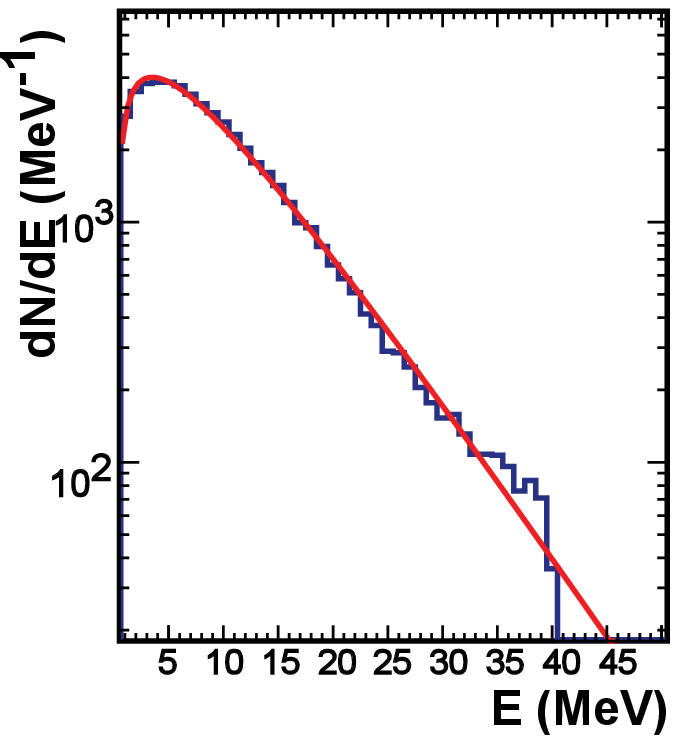}
\includegraphics[width=0.43\textwidth]{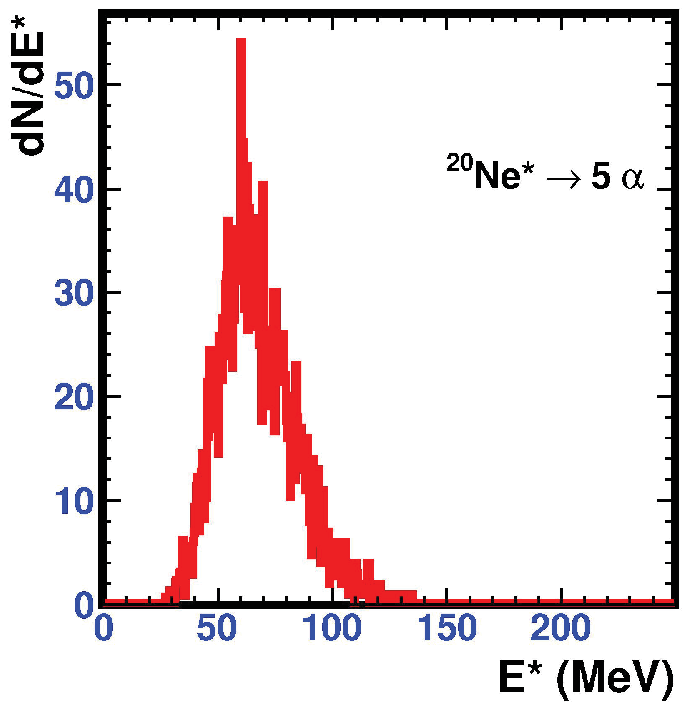}
\includegraphics[width=0.43\textwidth]{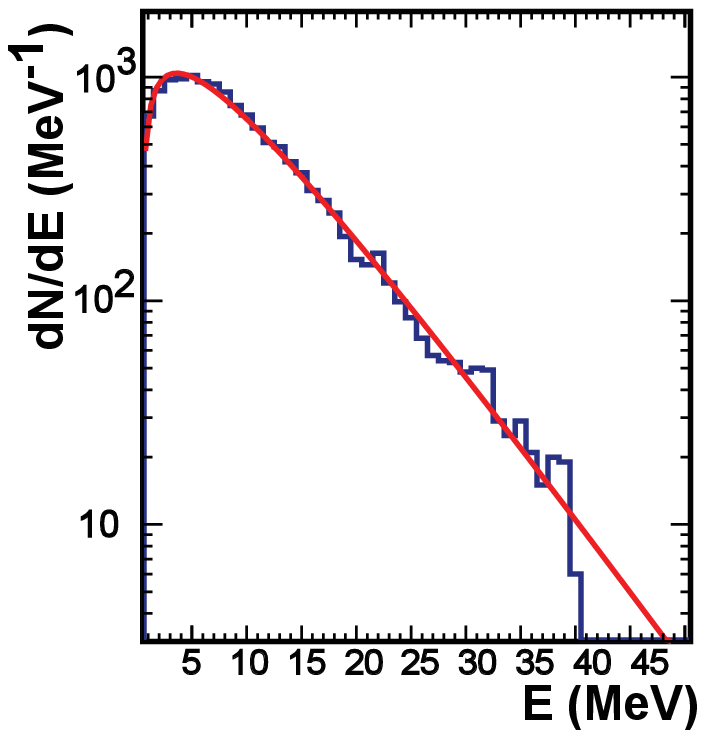}
\includegraphics[width=0.43\textwidth]{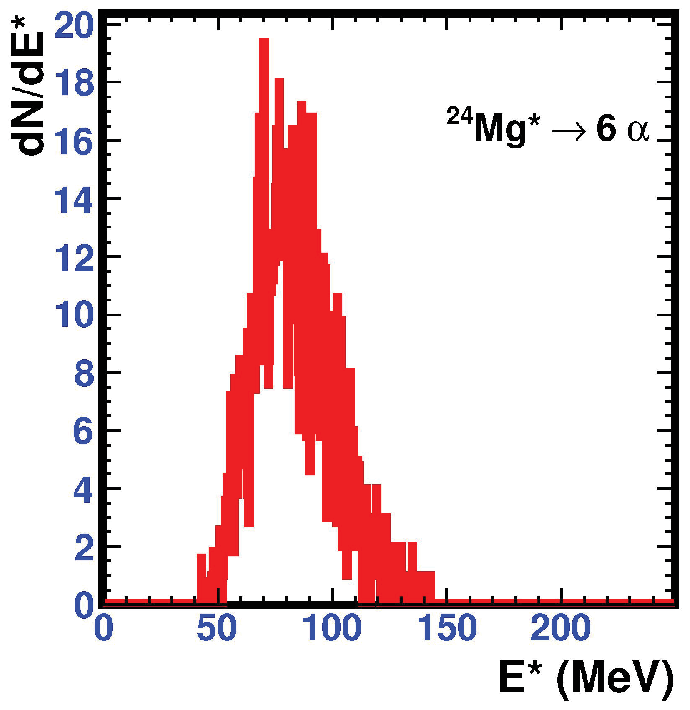}
\includegraphics[width=0.43\textwidth]{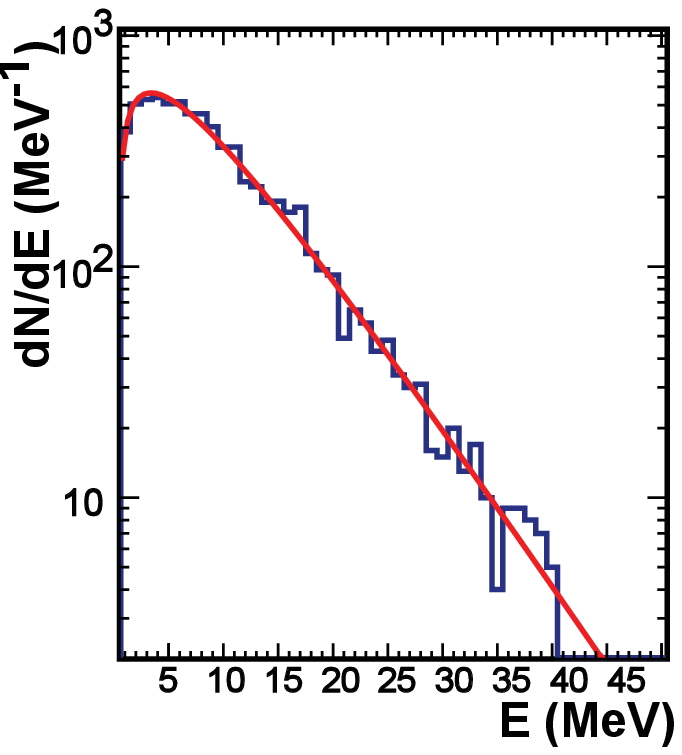}
\caption{\textit{Cont.}}
\label{fig:EXCIT_T_a}
\end{figure}

\addtocounter{figure}{-1}
\begin{figure}[!ht]
\centering
\includegraphics[width=0.43\textwidth]{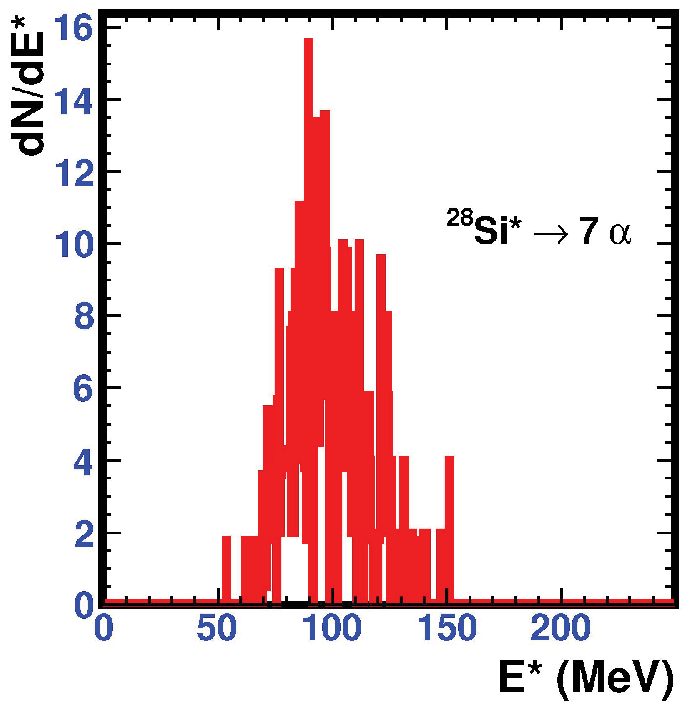}
\includegraphics[width=0.43\textwidth]{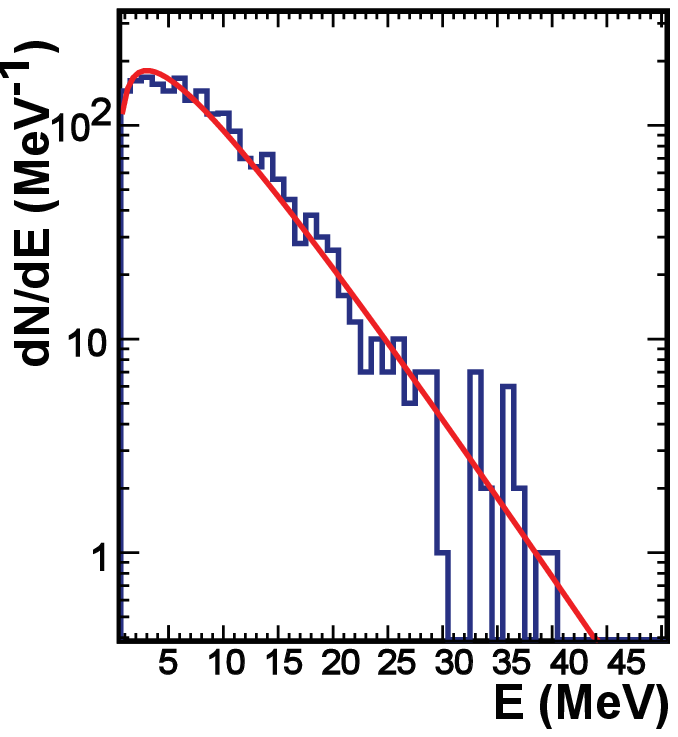}
\caption{
Excitation energy distributions (left hand side) and $\alpha$-particle energy
spectra (right hand side) for self-conjugate nuclei, from $^{16}O$ to
$^{28}Si$. Full curves superimposed on energy spectra are the results of 
Maxwellian fits (see text).\label{fig:EXCIT_T_b}}
\end{figure}
\begin{figure}[!ht]
\centering
\resizebox{0.5\textwidth}{!}{%
   \includegraphics{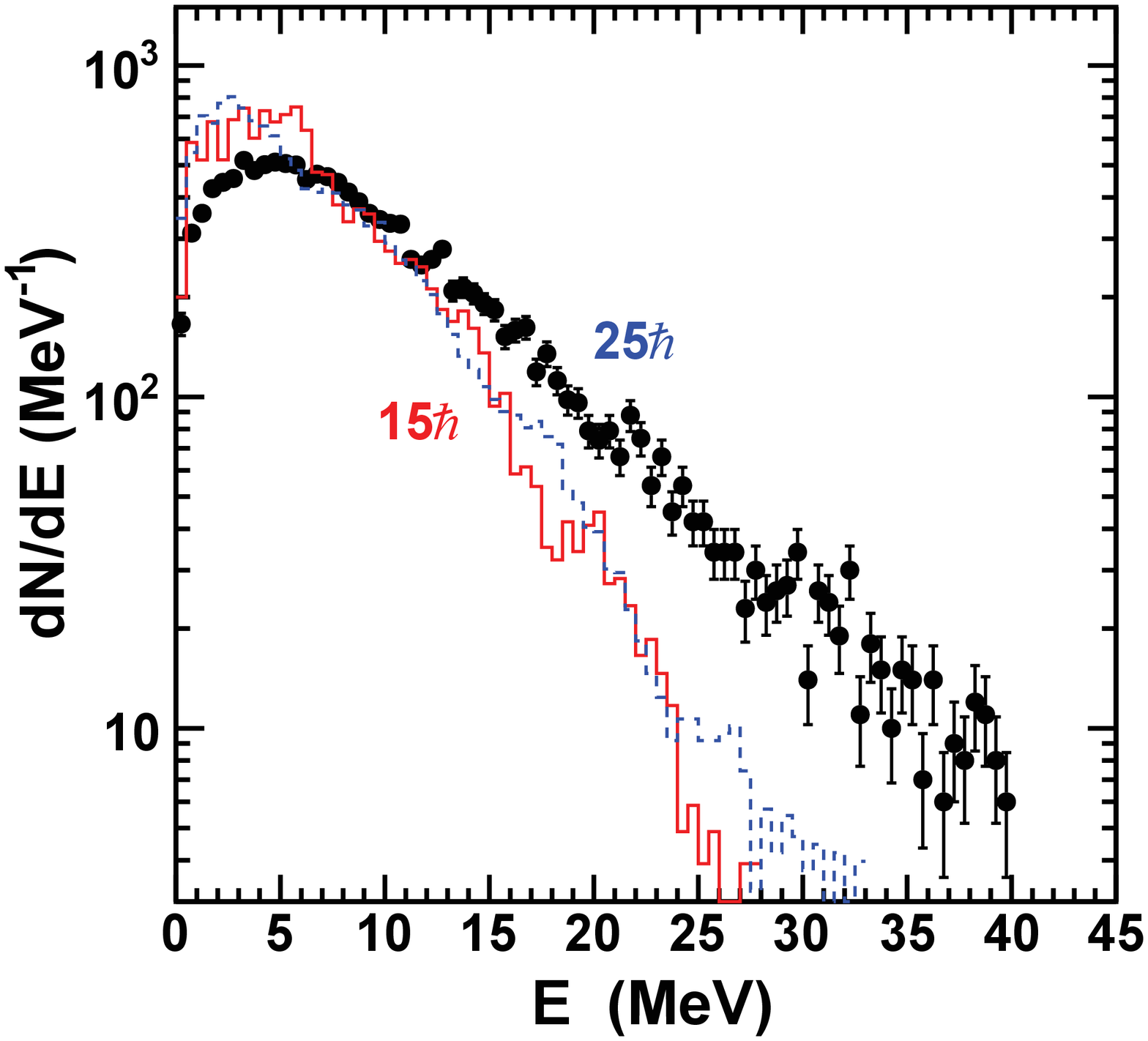}
   } 
	\caption{Sequential decay of excited $Ca$ projectiles: 
	energy spectra (in the $N\alpha$=5 system reference frame) of
	evaporated $\alpha$-particles associated to a  $^{20}Ne$
	evaporation residue . Full points are experimental data and
	histograms are results of GEMINI simulations (see text). 
   }
	\label{fig:SEQPF}
\end{figure}
\begin{figure}[!ht]
\centering
\resizebox{0.5\textwidth}{!}{%
   \includegraphics{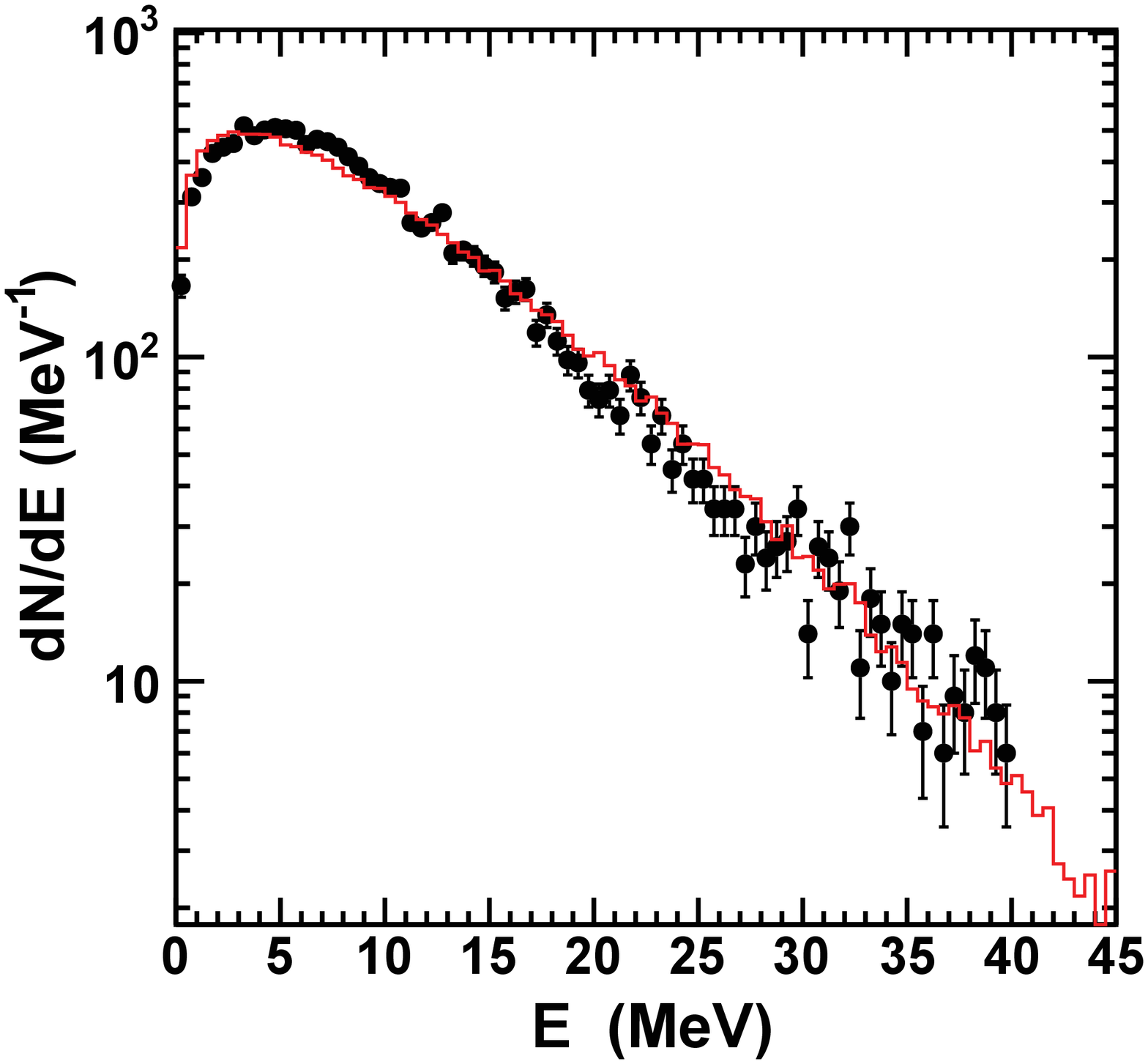}
   }
	\caption{Alpha particle spectrum from the $N\alpha$=5 system 
	($^{20}$Ne*). Black dots with
	statistical error bars correspond to experimental data (same as in
	Fig.2). Histogram 
	superimposed on data corresponds to filtered simulation of a
	simultaneous decay process (see text). 
    }
	\label{fig:EA5simul}
\end{figure}
 
Are $\alpha$-particles emitted sequentially or simultaneously?
To answer the question $\alpha$-energy spectra have been compared to
simulations. For excited Ca projectiles and $N\alpha$ sources, 
experimental velocity and excitation
energy distributions as well as distributions for spins were used as
inputs. The results of simulations were then
filtered using a software replica of the multi-detector including 
all detection and identification details. 
The simulated spectra are normalized to the area of experimental spectra. 
For sequential emission the GEMINI++ code~[\cite{Cha10}] was used.

Before discussing decays of $N\alpha$ sources, the possible evaporation 
from Ca projectiles, as stated previously, was considered.
Excitation energy for projectiles is deduced from 
$E^*$=$E^{*}(N\alpha)$+$E_{rel}$+$Q$.
$E_{rel}$ is the relative energy between the $N\alpha$ source and the
associated fragment (evaporation residue).
An example of the comparison between simulation and experimental
energy spectra of $\alpha$-particles is 
displayed in Figure~\ref{fig:SEQPF} (from~\cite{Bor15}); see also
Figure 6 of Ref.~\cite{Bor16} for $N\alpha$ 4 and 6.
They show a rather poor agreement indicating that the hypothesis 
of sequential evaporation of alpha particles is not correct.
Note that no more $^{24}$Mg, $^{20}$Ne or $^{16}$O evaporation residues
associated to $N\alpha$ from 4 to 7 are
produced in simulations for $^{40}Ca$ spin distributions
centered at values larger than 25$\hbar$.

Now, considering now sequential deexcitation of $N\alpha$ sources it appears,
as shown in Figure 5 of Ref.~\cite{Bor16},
that the agreement between 
data and simulations becomes poorer and poorer when $N\alpha$ value 
decreases. Moreover an important disagreement
between data and simulations
is observed for the percentages of $N\alpha$ sources which
de-excite via $^8Be$ emission~[\cite{Bor16}].

For simultaneous emission from $N\alpha$ sources, a dedicated
simulation was made that
mimics a situation in which $\alpha$ clusters are formed early
on while the
$N\alpha$ source is expanding~[\cite{girod_prl2013,ebran_2014}] due to thermal
pressure.
By respecting the experimental excitation energy distributions of 
$N\alpha$ sources shown in Figure~\ref{fig:EXCIT_T_b}, a distribution of 
$N\alpha$ events is generated as starting point of the simulation.
Event by event, the $N\alpha$ source is first split into $\alpha$'s.
Then the remaining available energy ($E^* + Q$) is directly randomly
shared among the $\alpha$-particles such as to conserve energy and linear
momentum~[\cite{Lop89}].
As an example, the histogram in Figure~\ref{fig:EA5simul}
(from~\cite{Bor15}) is the result of such
a simulation for $N\alpha$ = 5 ($^{20}$Ne*), 
which shows a good agreement with data; see also Figure 5 of Ref.~\cite{Bor16} and
Figure 2 of Ref.~\cite{Bor17} for $N\alpha$=4, 6, and 7.
Similar calculated energy spectra 
were also obtained with simulations containing an intermediate freeze-out
volume stage where $\alpha$-particles are formed and then propagated
in their mutual Coulomb field. This type of simulation is used to derive
density information from freeze-out volumes (see next subsection).
Note that $^8Be$ emission is out of the scope of the present
simulations. 

From these comparisons between both sequential and
simultaneous decay simulations it clearly appeared that sequential
emission was not able to reproduce experimental data whereas a remarkable
agreement is obtained when an
$\alpha$-clustering scenario is assumed. 
However one cannot exclude that a few
percent of $N\alpha$ sources, those produced with lower excitation
energies, sequentially de-excite. 

\subsection{Alpha clustering: temperature and density information}
Heavy-ion reactions at energies around the Fermi energy have
been shown good tools to create expanding nuclei and, more generally,
low density nuclear matter. It is why, in the past years, various
methods were used or developed to measure temperature and density in
such a nuclear matter~[\cite{Qin12,Pais20JPG,Mar16,Mab16}]. As we will
see in what follows, we are in our study in a simple case to extract
temperature and density information.

The excitation energy distributions of selected excited self-conjugate
nuclei ($N\alpha$ sources) and the corresponding kinetic energy spectra in their 
reference frame are displayed in Figure~\ref{fig:EXCIT_T_b}. 
Excitation energy thresholds for total deexcitation into $\alpha$-particles 
vary from 20 to 60 MeV when $N\alpha$ moves from 4 to 7 whereas mean 
excitation energy per nucleon is rather constant around 3.3-3.5 MeV.
Note that for the excitation spectrum of $^{28}$Si*, even if the
statistics are
limited, two peaks around 110 and 125 MeV could be present which were also observed
in Ref.~\cite{Cao19} and possibly related to toroidal high-spin isomers.
Kinetic energy spectra exhibit a thermal Maxwellian shape. We recall
that events with emitted preequilibrium $\alpha$-particles were
removed in our event selection (see Figure 3 of Ref.~\cite{Bor16}), which
explains our kinetic energy limit of 40 MeV.
\begin{table}[!h]
\centering 
\caption{Alpha-clustering for self-conjugate nuclei. 
Excitation energy information: mean value $<E^*>$ and standard
deviation $\sigma_{E^*}$.
Parameters from Maxwellian fits to energy spectra: temperature $T$ 
and Coulomb correction $C_c$. Densities normalized to saturation 
density $\rho_0$ have been deduced from simulations (see text). 
Statistical errors are within parentheses.}
\begin{tabular}{cccccc}
\hline
\textbf{nucleus} & \textbf{$<E^*>$(MeV)} &
\textbf{$\sigma_{E^*}$(MeV)} & \textbf{$T$(MeV)} &
     \textbf{$C_c$(MeV)} & \textbf{$\rho$/$\rho_0$}\\
$^{16}$O  & 52.4 & 15.7 & 6.15 (0.03) & 0.33 (0.03) & 0.37 (0.04)\\
$^{20}$Ne & 67.3 & 16.7 & 6.22 (0.05) & 0.45 (0.05) & 0.36 (0.04)\\				 
$^{24}$Mg & 83.5 & 17.4 & 5.92 (0.07) & 0.40 (0.07) & 0.34 (0.06)\\
$^{28}$Si & 98.5 & 17.6 & 5.40 (0.12) & 0.37 (0.16) & 0.34 (0.11)\\
\hline
\end{tabular}
\label{table1}
\end{table}
From these spectra, temperature, $T$, and the Coulomb
correction, $C_c$,  are extracted from a fit procedure using a thermal
Maxwellian formula: 
\[
dN/dE \propto (E- C_c)^{1/2} exp [-(E-C_c)/T],
\] 
with a volume pre-exponential factor, which is used for simultaneous
break-up such as alpha clustering~[\cite{Gol78}].
Curves in Figure~\ref{fig:EXCIT_T_b} correspond to Maxwellian fits
and Table~\ref{table1} summarizes all the results with statistical error bars.
Temperatures in the range of 5.3 - 6.3 MeV are extracted.
$C_c$ values which are deduced are low, around 0.3 - 0.5 MeV, which
qualitatively indicates rather low densities for expanding excited
nuclei at freeze-out due to thermal pressure.
Note that with a surface pre-exponential factor $(E - B_c)$ , which is used 
for sequential emissions, the best fits to the same data are
of lower quality and can only be achieved by allowing negative 
values for the Coulomb barrier $B_c$. Such
negative values have no physical meaning and the necessity to use a
volume pre-exponential factor further confirms the simultaneous emission
of $\alpha$-particles which characterizes the clustering.

To derive quantitative density information, the  Coulomb 
corrections parameters deduced from the fits, $C_c$,
have been used. As mentioned in the previous subsection dedicated
simulations were made to be compared with kinetic energy spectra. By
imposing temperatures deduced from the fits (see Table~\ref{table1})
within a simulation containing an intermediate freeze-out volume
stage it was easy to determine freeze-out volumes by also imposing 
an agreement with the most probable value of kinetic energy spectra
which is equal to T/2 + $C_c$. Freeze-out volumes values are in 
the range 2.7 - 3.0$V_0$, $V_0$ being the volumes of self-conjugate
nuclei at the saturation density. The corresponding normalized 
densities $\rho$/$\rho_0$ are indicated in Table~\ref{table1}.
\section{Discussion}
As mentioned in the introduction, two self-consistent mean field
calculations have been performed by imposing constrained deformation
with a restriction to spherical symmetric 
configurations~[\cite{girod_prl2013,ebran_2014}]. By gradually
increasing the nuclear radius, the density of self-conjugate nuclei is
decreased, and at a certain density value, the formation of clusters
appears. The first calculation was made with the constrained
Hartree-Fock-Bogoliubov approach using the Gogny D1S interaction, and in 
the second, the self-consistent relativistic
Hartree-Bogoliubov model with the effective interaction DD-ME2 has
been employed. 
For a comparison of density needed for $\alpha$-clustering one can
refer in calculations to the clustering radius normalized
to the ground state radius $r_c$/$r_{g.s.}$. 
In the first simulation this ratio is found around
1.8 for $^{16}O$ and $^{24}Mg$, whereas in the second, a smaller value
around 1.3 is obtained for $^{16}O$. The explanation given
in Ref.~\cite{ebran_2014} is the fact that the single-nucleon localization
is more pronounced with the relativistic functional, which facilitates
the formation of $\alpha$ clusters in excited states. Directly
translated into densities, these ratios correspond to $\rho$/$\rho_0$
around 0.17~[\cite{girod_prl2013}] and 0.45~[\cite{ebran_2014}]. 
However, it is important to stress that
calculations contain their own spurious center of mass energy which
should be removed. In Ref.~\cite{girod_prl2013} an estimate of the
correction needed was made starting from a calculation performed on
$^{8}Be$ constraining the distance between the two nascent
$\alpha$-particles. About 14 MeV were found to be missing 
to get twice the
binding energy of a single particle in the asymptotic limit. Thus, a
correction of 7 MeV per $\alpha$-particle was made, whatever the
self-conjugate nucleus (from $^{16}O$ to $^{28}Si$) and as a 
consequence the density for the appearance of $\alpha$-clustering was 
increased to $\rho$/$\rho_0$ $\sim$ 1/3, which is close to our
results. However, regarding these 
calculations, an important point one must underline is the fact that they 
correspond to zero temperature.  

In Ref.~\cite{Typ10} a generalized relativistic mean field model is used to
calculate $\alpha$-particle fractions in symmetric matter at
subsaturation density for different temperatures. In particular this
work describes the sudden decrease of the $\alpha$-particle fraction at high
densities by the vanishing of $\alpha$-particle binding energy due to 
the Pauli blocking that leads to the Mott effect, i.e., the dissolution of 
$\alpha$-particles in the medium. Thus, the maximum cluster
abundance is
reached around the Mott density. At a temperature of 6 MeV
this density is also $\sim$ $\rho_0$/3.

To conclude we can say that $\alpha$-clustering in self-conjugate
nuclei, experimentally deduced from $^{16}O$ to $^{28}Si$, occurs
around T = 5.5-6.0 MeV and densities in the range $\rho$/$\rho_0$ =
0.3-0.4. At present, constrained self consistent mean field
calculations at zero temperature are in qualitative agreement for
densities but even more realistic calculations are needed for a
valuable comparison with data.


\end{document}